# If the charged particle transport phenomenon in electrical circuits or in biological neurons is described with functions which satisfy differential equations using Riemann-Liouville or Caputo fractional order derivatives, then objectivity is lost.

Agneta M. Balint[1] and Stefan Balint[2]

[1]*Department of Physics, West University of Timisoara, Bulv. V. Parvan 4, 300223 Timisoara, Romania*

[2]*Department of Computer Science, West University of Timisoara, Bulv. V. Parvan 4, 300223 Timisoara, Romania*

Corresponding author: agneta.balint@e-uvt.ro; stefan.balint@e-uvt.ro

**Abstract.** In this paper the objectivity of the description of charged particles transport in electrical circuits, in biological neurons and biological neuron-networks is discussed. It is shown that the use of Riemann-Liouville or Caputo fractional order derivatives leads to the loose of the description objectivity.

**1. Objectivity in science and in the classical description of the electron transport phenomena in electrical circuits.** Objectivity in science means that qualitative and quantitative descriptions of phenomena remain unchanged when the phenomena are observed by different observers; that is, it is possible to reconcile observations of the process into a single coherent description of it [1]. In the following, the objectivity of the classical description of the electron transport in series RLC electrical circuit Fig.1. (resistor, inductor, capacitor) [2], is illustrated.

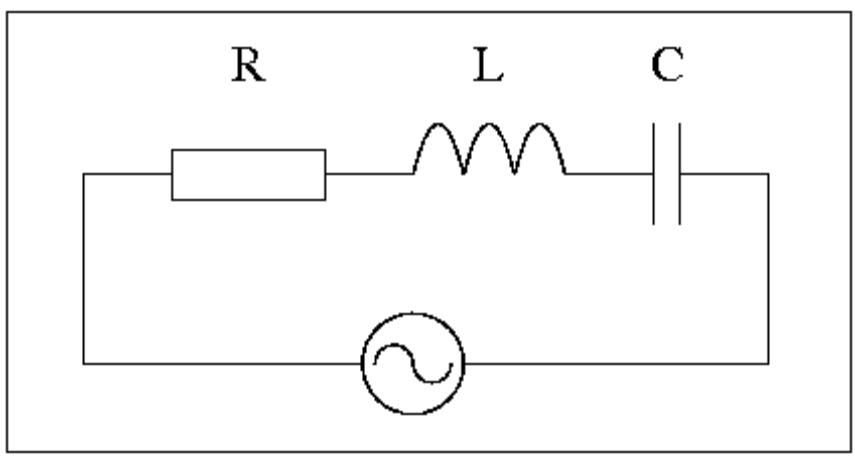

Fig.1. Classical series RLC electrical circuit.

One observer $O$ describes the electron transport in a classical series RLC electrical circuit presented in Fig.1. with a real valued and real variable function $i = i(t_M)$ representing the variation of the electron flow intensity (current intensity) in the circuit. More exactly, the observer $O$ fixes a moment of time $M_O$ for fixing the origin of the time measurement (for instant the moment of the start of his chronometer) and a unit [second] for the time measuring. A moment of time $M$ which is earlier than $M_O$ is represented by a negative real number $t_M < 0$, a moment of time $M$ which is later than $M_O$ is represented by a positive real number $t_M > 0$ and the moment of time $M_O$ is represented by the real number $t_{M_O} = 0$. At any moment of time $M$, represented by the number $t_M$, the observer measures the current intensity $i(t_M)$ and describes the transport with the function $i = i(t_M)$.

A second observer $O^*$ uses a similar procedure. For the observer $O^*$ the origin of the time measurement is $M_{O^*}$, the unit is [second]. A moment of time $M$ which is earlier than $M_{O^*}$ for the observer $O^*$ is represented by a negative number $t^*_M < 0$, the moment of time $M_{O^*}$ is represented by the real number $t^*_{M_{O^*}} = 0$ and a moment of time $M$ which is later than the moment of time $M_{O^*}$ is represented by a positive real number $t^*_M > 0$. At any moment of time $M$, represented by $t^*_M$, the observer $O^*$ measures the current intensity $i^*(t^*_M)$ and describes the transport with the function $i^* = i^*(t^*_M)$.

Remark that in case of the observer $O$, a moment of time $M$ is described by the real number $t_M$ and in case of the observer $O^*$ by the real number $t^*_M$. For the numbers $t_M$ and $t^*_M$ the following relations hold:

$$t_M = t^*_M + t_{M_{O^*}} \tag{1}$$

$$t^*_M = t_M + t^*_{M_O} \tag{2}$$

In the above relations $t_{M_{O^*}}$ is the real number which represents the moment $M_{O^*}$ in the system of time measuring of the observer $O$ and $t^*_{M_O}$ is the real number which represents the moment $M_O$ in the system of time measuring of the observer $O^*$.

At an arbitrary moment of time $M$, $i(t_M)$ represents the current intensity in the series electrical circuit RLC measured by the observer $O$ and $i^*(t^*_M)$ represents the current intensity in the same series electrical RLC circuit measured by the observer $O^*$. Because the current intensity concerns the same circuit the following relations hold:

$$i(t_M) = i^*(t^*_M) = i(t^*_M + t_{M_{O^*}}) \tag{3}$$

$$i^*(t^*_M) = i(t_M) = i^*(t_M - t^*_{M_{O^*}}) \tag{4}$$

Relations (3) or (4) reconcile the description made by the two observers, and make possible the description of the electron transport in a classical series electrical circuit RLC by one of the functions $i = i(t_M)$ or $i^* = i^*(t^*_M)$.

By using Ohm's law, Faraday's law, Kirchhoff's current law and Kirchhoff's voltage law it can be shown [3] that the current intensity variation described by the observer $O$ with the function $i = i(t_M)$ satisfies the following second order differential equation:

$$L \cdot \frac{d^2 i}{dt_M^2} + R \cdot \frac{di}{dt_M} + \frac{1}{C} i = \frac{dV}{dt_M} \tag{5}$$

Here: L- inductance, R- resistance, C- capacitance and $V = V(t_M)$ describes the variation of the external source voltage in terms of the observer $O$.

In terms of the description of observer $O*$, the same laws lead to the conclusion that the function $i*(t*_M)$ satisfies the following second order differential equation:

$$L \cdot \frac{d^2 i*}{dt*_M{}^2} + R \cdot \frac{di*}{dt*_M} + \frac{1}{C} i* = \frac{dV*}{dt*_M} \tag{6}$$

Here: L, R, C are the same constants as in (5) and $V* = V*(t*_M)$ describes the variation of the external source voltage in terms of the observer $O*$.

The descriptions $V = V(t_M)$ and $V* = V*(t*_M)$ are objective and so for them the following equalities hold:

$$V(t_M) = V*(t*_M) = V(t*_M + t_{M_{O*}}) = V*(t_M + t*_{M_O}) \tag{7}$$

Equations (5) and (6) are different but their solutions describe the current intensity variations in the same classical series electrical circuit RLC under the action of the same external source voltage. So it is necessary to show that the solutions of (5) and (6) verify (3) and (4). This can be proven showing that if $i = i(t_M)$ is a solution of (5), then $i* = i*(t*_M)$, defined by (3), is a solution of (6) and if $i* = i*(t*_M)$ is a solution of (6), then $i = i(t_M)$, defined by (4), is a solution of (5).

In other words, the dynamics of the current intensity in the considered electrical circuit can be described by the equation (5) or by the equation (6). This means that the descriptions (5) and (6) are independent on the observer. Each of them can be considered the description of the dynamics of current intensity in the considered series RLC electrical circuit.

**Remark.** By using Ohm's law, Faraday's law, Curie - von Schweidler current law and Kirchhoff's voltage law it can be shown [4], [5] that if the events "the RLC circuit start" and "the observer $O$ chronometer start" are simultaneous, then the current intensity variation described by the observer $O$ with the function $i = i(t_M)$ satisfies the following second order differential equation:

$$\frac{d^2 i}{dt_M{}^2} = -\frac{R}{L} \cdot \frac{di}{dt_M} - \frac{1}{L \cdot C} \cdot i + \frac{V_0}{h} [(\frac{R}{L} + \frac{1}{L \cdot C}) \cdot t_M{}^2 + \alpha \cdot (\alpha + 1)] \cdot t_M{}^{-\alpha-2} + \frac{V_0}{L} \tag{8}$$

That is because, according to von Schweidler [5] $i(t_M) = I(t_M) + i_C(t_M)$ is the Curie current [4], where $i_C(t_M) = \frac{V_0}{h} \cdot t_M{}^{-\alpha}$ (with $0 < \alpha < 1$) and $I = I(t_M)$ satisfies the equation

$$L \cdot \frac{d^2 I}{dt_M{}^2} + R \cdot \frac{dI}{dt_M} + \frac{1}{C} \cdot I = \frac{dV}{dt_M}.$$

For $t_{M_{O*}} > 0$ the description of the electron transport with the equation (8) is objective if the function $i*(t_M{}^*) = i(t_{M_{O*}} + t*) = i(t_M)$ verifies the equation:

$$\frac{d^2 i^*}{d(t_M^*)^2} = -\frac{R}{L} \cdot \frac{di^*}{dt_M^*} - \frac{1}{L \cdot C} \cdot i^* + \frac{V_0}{h} \cdot [(\frac{R}{L} + \frac{1}{L \cdot C}) \cdot (t_M^* + t_{M_{O^*}})^2 + \alpha \cdot (\alpha + 1)] \cdot (t_M^* + t_{M_{O^*}})^{-\alpha-2} + \frac{V_0}{L}$$

It is easy to show that the above condition is fulfilled.

Therefore, the description of the electron transport with differential equation (8) is objective.

## 2. Caputo, Riemann-Liouville fractional order derivatives

The Caputo fractional order derivative was introduced by M. Caputo in 1967 [6]. According to [7] for an indefinitely differentiable function $f:[0,\infty) \to R$ the Caputo fractional derivative of order $\alpha > 0$, is defined by :

$$D_C^\alpha f(t) = \frac{1}{\Gamma(n-\alpha)} \cdot \int_0^t \frac{f^n(\tau)}{(t-\tau)^{\alpha+1-n}} d\tau \qquad (9)$$

where $\Gamma$ is the Euler gamma function, $n = [\alpha] + 1$ and $[\alpha]$ is the integer part of $\alpha$.

For an indefinitely differentiable function $f:[0,\infty) \to R$ the Riemann-Liouville fractional derivative of order $\alpha > 0$, according to [5], is defined by :

$$D_{R-L}^\alpha f(t) = \frac{1}{\Gamma(n-\alpha)} \cdot \frac{d^n}{dt^n} \int_0^t \frac{f(\tau)}{(t-\tau)^{\alpha+1-n}} d\tau \qquad (10)$$

where $\Gamma$ is the Euler gamma function, $n = [\alpha] + 1$ and $[\alpha]$ is the integer part of $\alpha$.

If $\alpha > 0$ is integer, then the Caputo or Riemann-Liouville fractional order derivatives coincide with the classical integer order derivatives.

## 3. If in the description of the electron transport in a series RLC electrical circuit Caputo or Riemann-Liouville fractional order derivatives are used, then objectivity is lost.

In [8] the authors presents a classical derivation of fractional order circuits models. Generalized constitutive equations in terms of fractional order Riemann-Liouville derivatives are introduced in the Maxwell's equations. The Kirchhoff voltage law is applied in a RCL configuration and a fractional order differential equation is obtained for RCL circuit with Caputo derivatives.

In the following it is shown that the description of electron transport phenomenon in a series RLC circuit by substituting in (5) and (6) integer order derivatives of the current intensity with Caputo or Riemann-Liouville fractional order derivatives and the constants R, L, C with other constants depending on the order of derivatives, is not objective.

Consider the case $1 < \alpha < 2$, $0 < \beta < 1$, and the substitution of the integer order derivatives of the current intensity with Caputo fractional order derivative.

After the substitution for the observers $O$ and $O^*$ equations (5) and (6) become:

$$L_\alpha \cdot D_C^\alpha i(t_M) + R_\beta \cdot D_C^\beta i(t_M) + \frac{1}{C_{\alpha\beta}} \cdot i(t_M) = \frac{dV}{dt_M} \qquad (11)$$

$$L_\alpha \cdot D_C^\alpha i^*(t_M^*) + R_\beta \cdot D_C^\beta i^*(t_M) + \frac{1}{C_{\alpha\beta}} \cdot i^*(t_M^*) = \frac{dV^*}{dt^*_M} \qquad (12)$$

Assuming that $t_M > t_{M_{o^*}} > 0$ equalities:

$$D_C^\alpha i(t_M) = D_C^\alpha i^*(t^*_M) + \frac{1}{\Gamma(2-\alpha)} \cdot \int_0^{t_{M_{o^*}}} \frac{i''(\tau)}{(t_M - \tau)^{\alpha-1}} d\tau$$

$$D_C^\beta i(t_M) = D_C^\beta i^*(t^*_M) + \frac{1}{\Gamma(1-\beta)} \cdot \int_0^{t_{M_{o^*}}} \frac{i'(\tau)}{(t_M - \tau)^{\beta}} d\tau$$

implies that objectivity holds if and only if the following condition is satisfied:

$$\frac{L_\alpha}{\Gamma(2-\alpha)} \cdot \int_0^{t_{M_{o^*}}} \frac{i''(\tau)}{(t_M - \tau)^{\alpha-1}} d\tau + \frac{R_\beta}{\Gamma(1-\beta)} \cdot \int_0^{t_{M_{o^*}}} \frac{i'(\tau)}{(t_M - \tau)^{\beta}} d\tau = 0. \qquad (13)$$

Condition (13) in general is not satisfied.

The case $1 < \alpha < 2$, $0 < \beta < 1$, the substitution of the integer order derivatives of the current intensity with Riemann-Liouville fractional order derivative leads to the following objectivity condition:

$$\frac{L_\alpha}{\Gamma(2-\alpha)} \cdot \frac{d^2}{dt_M^2} \int_0^{t_{M_{o^*}}} \frac{i(\tau)}{(t_M - \tau)^{\alpha-1}} d\tau + \frac{R_\beta}{\Gamma(1-\beta)} \cdot \frac{d}{dt_M} \int_0^{t_{M_{o^*}}} \frac{i(\tau)}{(t_M - \tau)^{\beta}} d\tau = 0. \qquad (14)$$

Condition (14) in general is not satisfied.

**4. If, in the description of the ion transport through a passive membrane of a biological neuron cell Caputo or Riemann-Liouville fractional order derivatives are used, then objectivity is lost.**

In [9] Weinberg presents the passive cell membrane as a parallel resistor-capacitor circuit, assuming ideal capacitive behavior.

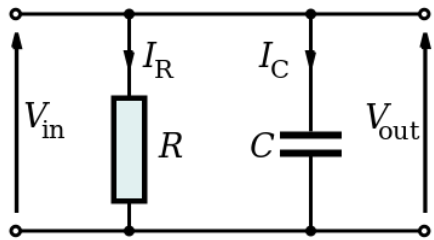

Fig.2. Parallel resistor-capacitor circuit

Under this hypothesis, the membrane voltages $V_m$ and $V_m^*$, used by observers $O$ and $O^*$ for the description of the membrane voltage dynamics, satisfy the equations:

$$C_m \cdot \frac{dV_m}{dt_M} + \frac{V_m}{R_m} = I(t_M)_{appl} \qquad (15)$$

$$C_m \cdot \frac{dV_m^*}{dt_M^*} + \frac{V_m^*}{R_m} = I*(t^*_M)_{appl} \qquad (16)$$

where: $C_m$ is the membrane capacitance, $R_m$ is the membrane resistance, $I(t_M)_{appl}$ and $I*(t^*_M)_{appl}$ represents the applied stimulus current in the observers $O$ and $O*$ descriptions, respectively. The applied stimulus current is objective so for the descriptions $I(t_M)_{appl}$ and $I*(t^*_M)_{appl}$ the following equality holds $I*(t^*_M)_{appl} = I(t_M)_{appl} = I(t^*_M + t_{M_{O^*}})_{appl}$.

Equations (15) and (16) are different, but their solutions describe the same membrane voltage dynamics, under the action of the same external stimulus current. So it is necessary to show that the solutions of (15) and (16) verify the objectivity condition:

$$V_m*(t_M*) = V_m(t_M) = V_m(t_M* + t_{M_{o^*}})$$

This can be made showing that, if $V_m = V_m(t_M)$ is a solution of (15), then $V_m* = V_m*(t^*_M)$ defined by $V_m*(t_M*) = V_m(t_M* + t_{M_{o^*}})$ is a solution of (16) and if $V_m* = V_m*(t^*_M)$ is a solution of (16), then $V_m(t_M) = V*_m(t_M - t_{M_{O^*}})$ is a solution of (15).

This means that, the dynamics of the membrane voltage can be described by the equation (15) or by the equation (16). In other words, the descriptions (15) and (16) are independent on the observer. Each of them can be used, for the description of the membrane voltage dynamics.

**Remark.** Also in [9], the author following some ideas presented in [4], [10], [11], [12], undertakes a study of the case when the description of the dynamics of a passive membrane satisfies a fractional order differential equation. The author of [9] underlines that: "the physiological source of such non-ideal capacitive behavior is not known, but we can speculate that this may arise due to heterogeneities in the dielectric properties of the membrane". The problem of objectivity of such type of description is not considered. In the following we show that: if the description of the ion transport through a passive membrane of a biological neuron cell satisfies Caputo or Riemann-Liouville fractional order differential equation, then objectivity is lost.

In case of Caputo fractional order differential equation, the membrane voltages descriptions $V_m$ and $V_m*$ used by observers $O$ and $O*$, satisfy the equations:

$$C_\alpha \cdot D_C^\alpha V_m(t_M) + \frac{1}{R_\alpha} \cdot V_m(t_M) = I(t_M)_{appl} \qquad (17)$$

$$C_\alpha \cdot D_C^\alpha V_m^*(t_M^*) + \frac{1}{R_\alpha} \cdot V_m^*(t_M^*) = I*(t^*_M)_{appl} \qquad (18)$$

In the above equations $t_M = t_M* + t_{M_{O^*}}$, $I*(t^*_M)_{appl} = I(t_M)_{appl} = I(t^*_M + t_{M_{O^*}})_{appl}$.

The objectivity condition is:

$$V_m*(t_M*) = V_m(t_M) = V_m(t_M*+t_{Mo*}) = V_m*(t_M - t_{Mo*}) \qquad (19)$$

Condition (19) is satisfied if and only if the next equality holds:

$$\frac{1}{\Gamma(1-\alpha)} \int_0^{t_{M)*}} \frac{V_m'(\tau)}{(t_M-\tau)^\alpha} d\tau = 0 \qquad (20)$$

In general the above condition is not fullfilled. So the description is not objective.

In case of the use of Riemann-Liouville fractional order derivatives, the membrane voltages $V_m$ and $V_m*$ used by observers $O$ and $O*$ for the description of dynamics satisfies the equations:

$$C_\alpha \cdot D_{R-L}^\alpha V_m(t_M) + \frac{1}{R_\alpha} \cdot V_m(t_M) = I(t_M)_{appl} \qquad (21)$$

$$C_\alpha \cdot D_{R-L}^\alpha V_m*(t_M*) + \frac{1}{R_\alpha} \cdot V_m*(t_M*) = I*(t*_M)_{appl} \qquad (22)$$

In the above equations $t_M = t_M*+t_{Mo*}$, $I*(t*_M)_{appl} = I(t_M)_{appl} = I(t*_M+t_{Mo*})_{appl}$.

The objectivity condition is:

$$V_m*(t_M*) = V_m(t_M) = V_m(t_M*+t_{Mo*}) = V_m*(t_M - t_{Mo*}) \qquad (23)$$

Condition (23) is satisfied if and only if the next equality holds:

$$\frac{1}{\Gamma(1-\alpha)} \cdot \frac{d}{dt} \int_0^{t_{M)*}} \frac{V_m(\tau)}{(t_M-\tau)^\alpha} d\tau = 0 \qquad (24)$$

In general the above condition is not fullfilled. So, the description is not objective.

**5. If in the Hodgkin-Huxley description of the ion transport through a membrane of a biological neuron cell, Caputo or Riemann-Liouville fractional order derivatives are used, then objectivity is lost.**

In terms of the observers $O$ and $O*$ the ions transport through the membrane of a biological cell in the classical Hodgkin-Huxley description [10] are the solutions of the following systems of differential equations:

$$C \cdot \frac{dV}{dt_M} = I(t_M) - g_{Na} \cdot m^3 \cdot h \cdot (V-E_{Na}) - g_K \cdot n^4 \cdot (V-E_K) - g_L \cdot (V-E_L)$$

$$\frac{dm}{dt} = \alpha_m \cdot (1-m) - \beta_m \cdot m$$

$$\frac{dh}{dt_M} = \alpha_h \cdot (1-h) - \beta_h \cdot h \qquad (25)$$

$$\frac{dn}{dt_M} = \alpha_n \cdot (1-n) - \beta_n \cdot n$$

$$C \cdot \frac{dV^*}{dt_M^*} = I^*(t_M^*) - g_{Na} \cdot m^{*3} \cdot h^* \cdot (V^* - E_{Na}) - g_K \cdot n^{*4} \cdot (V^* - E_K) - g_L \cdot (V^* - E_L)$$

$$\frac{dm^*}{dt_M^*} = \alpha_m \cdot (1 - m^*) - \beta_m \cdot m^*$$

$$\frac{dh^*}{dt_M^*} = \alpha_h \cdot (1 - h^*) - \beta_h \cdot h^* \tag{26}$$

$$\frac{dn^*}{dt_M^*} = \alpha_n \cdot (1 - n^*) - \beta_n \cdot n^*$$

where: $V = V_m - V_{rest}$ represents the membrane potential $V_m$ relative to the rest $V_{rest}$; $m$, $h$ and $n$ are the sodium current $I_{Na}$ activation, $I_{Na}$ inactivation and potassium current $I_K$ activation gating variables, respectively, $t_M = t_M^* + t_{M_{O^*}}$, $I(t_M) = I^*(t^*_M) = I(t^*_M + t_{M_{O^*}}) = I^*(t_M - t_{M_{O^*}})$ is a time dependent stimulus. For the meaning of the other symbols see [10].

It has to be noted that the objectivity requirement is satisfied if and only if the following equalities hold: $V^*(t_M^*) = V(t_M)$; $m^*(t_M^*) = m(t_M)$; $h^*(t_M^*) = h(t_M)$; $n^*(t_M^*) = n(t_M)$.

This can be made showing that, if $V(t_M)$, $m(t_M)$, $h(t_M)$, $n(t_M)$ is a solution of (25), then $V^*(t_M^*) = V(t_M^* + t_{M_{O^*}})$, $m^*(t_M^*) = m(t^*_M + t_{M_{O^*}})$, $h^*(t_M^*) = h(t^*_M + t_{M_{O^*}})$, $n^*(t_M^*) = n(t^*_M + t_{M_{O^*}})$ is a solution of (26) and if $V^*(t^*_M)$, $m^*(t_M^*), h^*(t_M^*), n^*(t_M^*)$ is a solution of (26), then $V(t_M) = V^*(t_M - t_{M_{O^*}})$, $m(t_M) = m^*(t_M - t_{M_{O^*}}), h(t_M) = h^*(t_M - t_{M_{O^*}})$, $n(t_M) = n^*(t_M - t_{M_{O^*}})$ is a solution of (25).

So, the classical Hodgkin-Huxley description is objective.

**Remark.** In [9], the author undertakes a study also of the case when the Hodgkin-Huxley description of the sodium and potassium ions transport are described by functions which satisfy system of Caputo fractional order differential equations. In [9] the objectivity problem for this type of description is not considered. In the following it is shown that: if the description of Hodgkin-Huxley type of sodium and potassium ions transport, through a membrane of a biological neuron cell, satisfies Caputo fractional order differential equations, then objectivity is lost.

In case of Caputo fractional order derivatives, for the two observers $O$ and $O^*$, the equations used in [9] are the following (fractional-order membrane patch model):

$$C^\alpha_m \cdot D_C^\alpha V(t_M) = I(t_M) - g_{Na} \cdot m^3 \cdot h \cdot (V - E_{Na}) - g_K \cdot n^4 \cdot (V - E_K) - g_L \cdot (V - E_L)$$

$$\frac{dm}{dt_M} = \alpha_m \cdot (1-m) - \beta_m \cdot m$$

$$\frac{dh}{dt_M} = \alpha_h \cdot (1-h) - \beta_h \cdot h \qquad (27)$$

$$\frac{dn}{dt_M} = \alpha_n \cdot (1-n) - \beta_n \cdot n$$

$$C^\alpha_m \cdot D_C^\alpha V^*(t_M^*) = I^*(t_M^*) - g_{Na} \cdot m^{*3} \cdot h^* \cdot (V^* - E_{Na}) - g_K \cdot n^{*4} \cdot (V^* - E_K) - g_L \cdot (V^* - E_L)$$

$$\frac{dm^*}{dt_M^*} = \alpha_m \cdot (1-m^*) - \beta_m \cdot m^*$$

$$\frac{dh^*}{dt_M^*} = \alpha_h \cdot (1-h^*) - \beta_h \cdot h^* \qquad (28)$$

$$\frac{dn^*}{dt_M^*} = \alpha_n \cdot (1-n^*) - \beta_n \cdot n^*$$

The objectivity requirement is satisfied if and only if the following equalities hold: $V^*(t_M^*) = V(t_M)$; $m^*(t_M^*) = m(t_M)$; $h^*(t_M^*) = h(t_M)$; $n^*(t_M^*) = n(t_M)$. These equalities hold if and only if the following equality holds:

$$\frac{1}{\Gamma(1-\alpha)} \int_0^{t_{M)^*}} \frac{V'(\tau)}{(t_M - \tau)^\alpha} d\tau = 0 \qquad (29)$$

In general condition (29) is not fullfilled. So, the fractional-order membrane patch model is not objective.

With similar arguments it can be shown that the ***Fractional order nerve axon model***, *the **Fractional order Hodgkin-Huxley model extended to a cable***, and the ***Fractional order neural network model presented*** in [9] are not objective. It can be shown similarly that the ***Electromechanical fractional order Hodgkin-Huxley model*** presented in [21] is also not objective.

## 6. Introducing fractional order derivative in the voltage oscillation description of the Barnacle Giant Muscle Fiber presented by Morris and Lecar, the description becomes non objective.

Morris and Lecar in [13] show that voltage-clamp studies of the barnacle muscle made in [14]- [18] indicate that the fiber possesses a simply conductance system consisting of voltage dependent $Ca^{++}$ and $K^+$ channels, neither of which innactivates appreciably. Current clamp studies [17] and [19], however, show complicated oscillatory voltage behavior. The mathematical study developed in [13] reveals that this simple system can predict much of the brancle fiber behavior, although the simplest model fails to explain some areas of behavior.

For the observers $O$ and $O^*$ the systems of differential equations considered in [13] are the followings:

$$C \cdot \frac{dV}{dt} = I(t) - g_{Ca} \cdot M \cdot [V - V_{Ca}] - g_K \cdot N \cdot [V - V_K] - g_L \cdot [V - V_L]$$

$$\frac{dM}{dt} = \lambda_M(V) \cdot [M_\infty(V) - M] \qquad (30)$$

$$\frac{dN}{dt} = \lambda_N(V) \cdot [N_\infty(V) - N]$$

$$C \cdot \frac{dV^*}{dt^*} = I^*(t^*) - g_{Ca} \cdot M \cdot [V^* - V_{Ca}] - g_K \cdot N^* \cdot [V^* - V_K] - g_L \cdot [V^* - V_L]$$

$$\frac{dM^*}{dt^*} = \lambda_M(V^*) \cdot [M_\infty(V^*) - M^*] \qquad (31)$$

$$\frac{dN^*}{dt^*} = \lambda_N(V^*) \cdot [N_\infty(V^*) - N^*]$$

The meaning of the symbols appearing in the above systems can be found in [13] and for the numbers representing the moment of time, the following notation $t = t_M$ and $t^* = t^*_M$ were used.

It is easy to show that the dynamics described by (30) is the same as that described by (31). In other words, the Morris-Lecar description is objective.

**Remark.** In the following it is shown that: if in the description of Morris-Lecar type of $Ca^{++}$ and $K^+$ ions transport, through the fiber, Caputo fractional order derivatives are used, then objectivity is lost.

In case of Caputo fractional order derivatives, for the two observers $O$ and $,O^*$ equations (30), (31) lead to the equations:

$$C_\alpha \cdot D_C^\alpha V(t) = I(t) - g_{Ca} \cdot M \cdot [V - V_{Ca}] - g_K \cdot N \cdot [V - V_K] - g_L \cdot [V - V_L]$$

$$D_C^\beta M(t) = \lambda_M(V) \cdot [M_\infty(V) - M] \qquad (32)$$

$$D_C^\gamma N(t) = \lambda_N(V) \cdot [N_\infty(V) - N]$$

$$C_\alpha \cdot D_C^\alpha V^*(t^*) = I^*(t^*) - g_{Ca} \cdot M^* \cdot [V^* - V_{Ca}] - g_K \cdot N^* \cdot [V^* - V_K] - g_L \cdot [V^* - V_L]$$

$$D_C^\beta M^*(t^*) = \lambda_M(V^*) \cdot [M_\infty(V^*) - M^*] \qquad (33)$$

$$D_C^\gamma N^*(t^*) = \lambda_N(V^*) \cdot [N_\infty(V^*) - N^*]$$

The objectivity requirement is satisfied if and only if the following equalities hold: $V^*(t_M^*) = V(t_M)$; $m^*(t_M^*) = m(t_M)$; $h^*(t_M^*) = h(t_M)$; $n^*(t_M^*) = n(t_M)$. The above equalities hold if and only if the following conditions are satisfied:

$$\frac{C_\alpha}{\Gamma(1-\alpha)} \cdot \int_0^{t_{M_{Q^*}}} \frac{V'(\tau)}{(t-\tau)^\alpha} d\tau = 0; \quad \frac{1}{\Gamma(1-\beta)} \cdot \int_0^{t_{M_{Q^*}}} \frac{N'(\tau)}{(t-\tau)^\beta} d\tau = 0 ;$$

$$\frac{1}{\Gamma(1-\gamma)} \cdot \int_0^{t_{M_{Q^*}}} \frac{M'(\tau)}{(t-\tau)^\gamma} d\tau = 0 \qquad (34)$$

In general the above conditions are not fullfilled. Therefore, the Morris –Lecar model considered in [20] is not objective.

## 7. Conclusion

The main conclusion of this study is in fact the following question: if a mathematical description of a real phenomenon is not objective, then what is the interpretation of the reported results and how these results have to be used?